\documentclass{ws-ijmpb}
\usepackage{color}
\usepackage{graphicx}
\usepackage{dcolumn}
\usepackage{bm}
\newcommand{\bra}[1]{\langle#1|}
\newcommand{\ket}[1]{|#1\rangle}

\providecommand{\openone}{\leavevmode\hbox{\small1\kern-3.8pt\normalsize1}}

\begin{document}

\markboth{R. Lo Franco, B. Bellomo, S. Maniscalco, G. Compagno}
{Dynamics of quantum correlations in two-qubit systems within non-Markovian environments}

%
\catchline{}{}{}{}{}
%

\title{DYNAMICS OF QUANTUM CORRELATIONS IN TWO-QUBIT SYSTEMS WITHIN NON-MARKOVIAN ENVIRONMENTS}

\author{ROSARIO LO FRANCO$^{\ast,\P}$, BRUNO BELLOMO$^{\dagger}$, SABRINA MANISCALCO$^{\ddagger,\S}$, and GIUSEPPE COMPAGNO$^{\ast,\Vert}$}
\address{$^\ast$CNISM and Dipartimento di Fisica, Universit\`a di Palermo, via Archirafi 36, 90123 Palermo, Italy\\
$^\dagger$Universit\'e Montpellier 2, Laboratoire Charles Coulomb UMR 5221, F-34095, Montpellier, France\\
$^\ddagger$Turku Center for Quantum Physics, Department of Physics and Astronomy, University of Turku, FIN20014, Turku, Finland\\
$^\S$SUPA, EPS Physics, Heriot-Watt University, Edinburgh, EH14 4AS, United Kingdom\\
$^\P$rosario.lofranco@unipa.it\\
$^\dagger$bruno.bellomo@univ-montp2.fr\\
$^\S$s.maniscalco@hw.ac.uk\\
$^\Vert$giuseppe.compagno@unipa.it}

\maketitle

\begin{history}
\received{\today}
\revised{Day Month Year}
\end{history}

\begin{abstract}
Knowledge of the dynamical behavior of correlations with no classical counterpart, like entanglement, nonlocal correlations and quantum discord, in open quantum systems is of primary interest because of the possibility to exploit these correlations for quantum information tasks. Here we review some of the most recent results on the dynamics of correlations in bipartite systems embedded in non-Markovian environments that, with their memory effects, influence in a relevant way the system dynamics and appear to be more fundamental than the Markovian ones for practical purposes. Firstly, we review the phenomenon of entanglement revivals in a two-qubit system for both independent environments and a common environment. We then consider the dynamics of quantum discord in non-Markovian dephasing channel and briefly discuss the occurrence of revivals of quantum correlations in classical environments.   
\end{abstract}


\section{Introduction}

Realistic quantum systems are important both for their fundamental role\cite{schrodinger} and for up-to-date technologies\cite{nielsenchuang}. They are open, in the sense that their correlations with surrounding environments are of statistical nature and unavoidable\cite{petru}. The characterization of correlations in composite open quantum systems is essential for quantum information and computation tasks\cite{benenti}. To a specific property of the system  it can be made to correspond a type of correlation that is used to define the property itself. In fact, entanglement represents correlations related to non-separability of the state of a composite quantum system; violation of Bell inequalities\cite{bell,clauser} represents quantum nonlocality, that is correlations not reproducible by any local classical model; quantum discord (entropic\cite{Zurek2001PRL,vedral2001JPA} and geometric\cite{dakic2010PRL}) represents quantumness and indicators of purely classical correlations can also be defined. Each of these types of correlations can be useful for specific tasks. For example, entanglement is the main resource for teleportation and superdense coding\cite{nielsenchuang,bennett}, nonlocal correlations guarantee secure quantum cryptography\cite{acin2006PRL,gisin2007natphoton}, entropic discord is a resource for deterministic one-qubit computation\cite{Knill1998PRL,Lanyon2008PRL,Datta2008PRL} and geometric discord for remote state preparation\cite{DakicZeilinger2012arXiv,adesso2012arxiv}. The above correlations, each linked to a specific property of bipartite quantum systems, have been extended to multipartite systems and quantified  in terms of distances between the state of the system and its closest states without that property\cite{dakic2010PRL,Modi2010PRL,luo2011PRL,bellomo2012PRA,bellomo2012linearentropy}. It is therefore essential to understand, in a given open system, the dynamics of the various correlations and in particular how much and for how long they persist.

The dynamical behavior of correlations present in a composite open quantum system strongly depend on the noise produced by the surrounding environment.  In this sense one of the most important aspects of the environment is if it can be described either as memoryless (Markovian) or as with memory (non-Markovian). The characterization of the environment as Markovian or non-Markovian, with respect to a given system, is determined by the ratio between its typical correlation time and the system relaxation time. For two-qubit systems in Markovian environments, some kind of correlations may be subject to early-stage disappearance (ESD, or sudden death), meaning that they completely disappear at a finite time in spite of exponential decay of single-qubit coherence. This has been shown, both theoretically and experimentally, in the case of entanglement\cite{diosi,yu1,santos,carv1,qasimi2008PRA,yu2009Science,almeida,kimble2007PRL} and of Bell inequality violations\cite{miran2004PLA,kofman2008PRA}. This phenomenon strongly limits the time when correlations can be usefully exploited. Differently, two-qubit systems in structured non-Markovian environments may present phenomena such as revivals\cite{bellomo2007PRL,bellomo2008PRA,Xu2010PRL,bellomo2011PhyScrSavasta,maniscalco2008PRL,mazzola2009PRA,bellomo2008bell,fanchini2010PRA,wang2010PRA,bellomo2011IJQI} and trapping\cite{bellomo2008trapping,bellomo2009ASL,bellomo2010PhysScrManiscalco} of correlations, thus overcoming ESD. The study of the dynamics of correlations under non-Markovian noise thus appears to be fundamental for quantum information purposes\cite{yu2009Science} and it shall be the main subject of this paper. 

In particular, the aim of this review is to summarize some of the most recent results on the dynamics of correlations in bipartite qubit systems embedded in specific non-Markovian environments. We shall first review the phenomenon of correlation revivals in a two-qubit system both for independent environments and for a common environment. Secondly, we describe how structured environments, as photonic crystals, may protect correlations against decoherence. We finally discuss the dynamics of quantum correlations in non-dissipative dephasing channels and in classical environments given by random external fields.

\section{Revivals of entanglement}
In this section we review the phenomenon of entanglement revivals for non-Markovian bosonic reservoirs, both in the case of independent environments and in a common environment. 

\subsection{Two-qubit initial states}
We first define the initial states typically chosen for the analysis and describe their properties. We consider initially entangled two-qubit states within the class of X states\cite{yu5}, whose general structure is, in the standard computational basis $\mathcal{B}=\{\ket{1}\equiv\ket{11},\ket{2}\equiv\ket{10}, \ket{3}\equiv\ket{01}, \ket{4}\equiv\ket{00} \}$,
\begin{equation}\label{Xstates}
   \hat{\rho}_X= \left(
\begin{array}{cccc}
  \rho_{11} & 0 & 0 & \rho_{14} \\
  0 & \rho_{22} & \rho_{23} & 0 \\
  0 & \rho_{23}^* & \rho_{33} & 0 \\
  \rho_{14}^* & 0 & 0 & \rho_{44}\\
\end{array}
\right).
\end{equation}
X-structured density matrices may arise in a wide variety of physical situations\cite{bose2001,Pratt2004PRL,peters2004PRL,wang,hagley1997PRL,chiuriPRA2011,dicarlo2009Nature}. These states are also encountered as eigenstates in all the systems with odd-even symmetry like in the Ising and the XY models\cite{Fazio2002Nature,osborne2002PRA}. Moreover, for many physical dynamics of open quantum systems an initial X structure is maintained in time\cite{yu5}. This shall occur, in particular, in the following cases described in this review. X states contain the class of extended Werner-like (EWL) states, defined as\cite{bellomo2008PRA}
\begin{equation}\label{EWL}
    \hat{\rho}^\Phi=r \ket{\Phi}\bra{\Phi}+(1-r)\openone/4,\quad
    \hat{\rho}^\Psi=r \ket{\Psi}\bra{\Psi}+(1-r)\openone/4.
\end{equation}
In Eq.~(\ref{EWL}), $r$ is the purity parameter, $\openone$ the $4\times4$ identity matrix and
\begin{eqnarray}
\ket{\Phi}=a\ket{01}+b\ket{10},\quad
\ket{\Psi}=a\ket{00}+b\ket{11},\label{Bell-likestates}
\end{eqnarray}
are the Bell-like states with $a$ real, $b=|b|e^{i\gamma}$ and $a^2+|b|^2=1$. For $a=\pm b=1/\sqrt{2}$, Bell-like states become the Bell states $\ket{\Phi^\pm}=\left(\ket{01}\pm\ket{10}\right)/\sqrt{2}$, $\ket{\Psi^\pm}=\left(\ket{00}\pm\ket{11}\right)/\sqrt{2}$. The states defined by Eq.~(\ref{EWL}) reduce to the well-known Werner-like states\cite{munro,wei} when their pure part is a Bell state. For $r=0$ EWL states become totally mixed states, while for $r=1$ they reduce to the Bell-like states $\ket{\Phi},\ket{\Psi}$ of Eq.~(\ref{Bell-likestates}). One of the important aspects of the EWL states $\hat{\rho}^\Phi(0),\hat{\rho}^\Psi(0)$ is that they allow to study the effect, on the dynamics of correlations, of both the initial state mixedness and the degree of entanglement of their pure part.

The entanglement of the bipartite system shall be hereafter quantified by the concurrence\cite{wootters}. For a X state, as defined
in Eq.~(\ref{Xstates}), concurrence can be easily computed and is given by\cite{yu5}
\begin{equation}\label{concurrence x state}
C_{\rho_X}=2\mathrm{max}\{0,K_1,K_2\},\
K_1=|\rho_{23}|-\sqrt{\rho_{11}\rho_{44}},\
K_2=|\rho_{14}|-\sqrt{\rho_{22}\rho_{33}}.
\end{equation}
EWL states of Eq.~(\ref{EWL}) give equal value for the concurrences, given by
\begin{equation}
C_\rho^{\Phi}=C_\rho^{\Psi}=2\mathrm{max}\{0,(a|b|+1/4)r-1/4\}.
\end{equation}
From previous equation one finds that there is entanglement when $r>r^\ast=(1+4a|b|)^{-1}$ (for $a=|b|=1/\sqrt{2}$ one has $r^\ast=1/3$). In particular, the Bell-like pure states $\ket{\Phi},\ket{\Psi}$ have the same degree of entanglement (concurrence) $C_\Phi(0)=C_\Psi(0)=2a\sqrt{1-a^2}$.

\subsection{Independent bosonic environments} \label{sec:inden}
\begin{figure}[bt]
{\centerline{\psfig{file=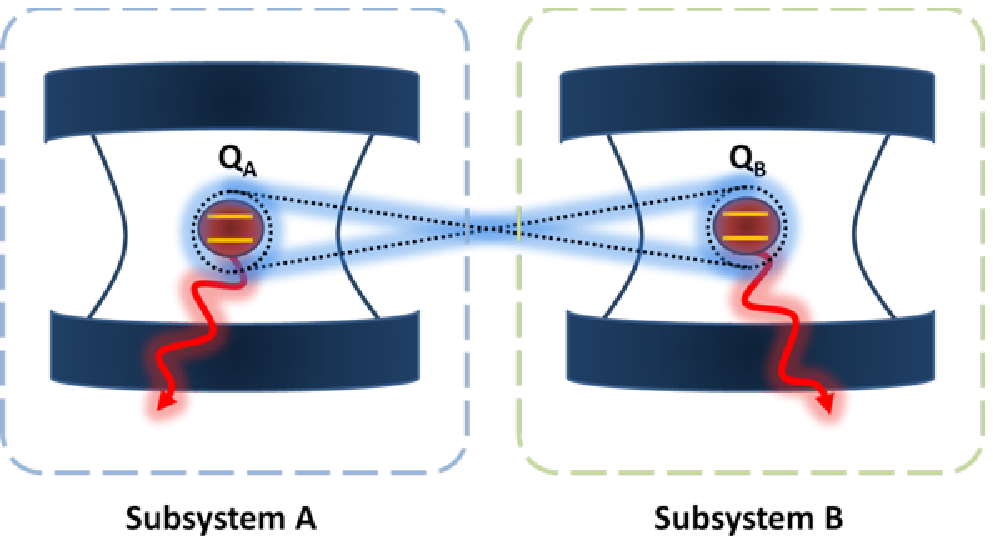,width=6 cm}\hspace{1.5 cm}
\psfig{file=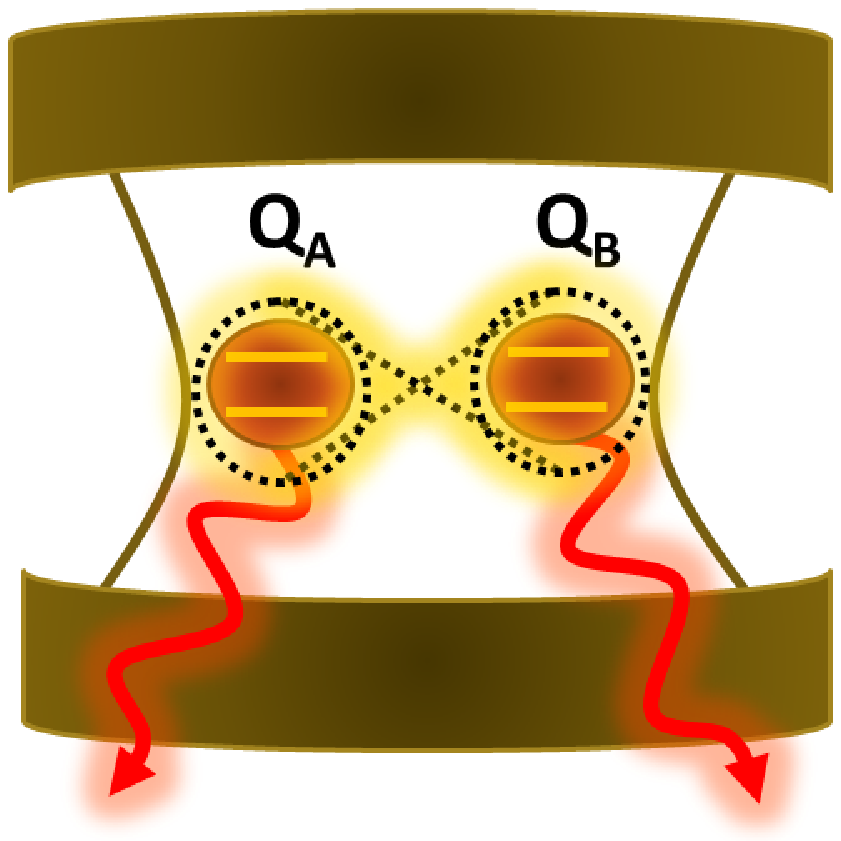,width=3.2 cm}}}
\vspace*{8pt}
\caption{\label{fig:system}Schematic of the two-qubit systems with independent environments (left figure) and common environment (right figure).}
\end{figure}

We consider a system composed by two independent parts each made of a qubit placed inside a zero-temperature bosonic environment (see left part of Fig.~\ref{fig:system}). The two qubits $A$ and $B$, with ground and excited states $\ket{0}$, $\ket{1}$ and transition frequency $\omega_0$, are identical and initially entangled. 
A procedure has been introduced, applicable to a general multipartite system made by independents subsystems, that allows to obtain the reduced density matrix of the system at the time $t$ by the knowledge of that of the single subsystems\cite{bellomo2007PRL,bellomo2008PRA}. Here we summarize this procedure for the case of a two-qubit system. For independent parts the total time evolution operator can be factorized in two terms relative to the two parts. Under these conditions, given the dynamics of each qubit as $\rho^{A}_{ii'}(t)=\sum_{ll'}A_{ii'}^{ll'}(t)\rho^{A}_{ll''}(0)$, $\rho^{B}_{jj'}(t)=\sum_{mm'}B_{jj'}^{mm'}(t)\rho^{B}_{mm'}(0)$, the two-qubit density matrix is simply given by\cite{bellomo2007PRL} 
\begin{equation}\label{totalevo}
\rho_{ii',jj'}(t)=\sum_{ll',mm'}A_{ii'}^{ll'}(t)B_{jj'}^{mm'}(t)  \rho^{}_{ll',mm'}(0),
\end{equation}
where the indexes $i,j,l,m=0,1$. This procedure is easily generalizable to an arbitrary multipartite system of qudits ($d$-level subsystems)\cite{bellomo2008PRA} and it can be also applied to the case of two qubits in a common environment provided that the qubits are sufficiently separated in order to neglect the dipole-dipole interaction and their distance is larger than the spatial correlation length of the reservoir.

The dynamics of each qubit coupled to the environment modes is described by the spin-boson Hamiltonian\cite{petru}
\begin{equation}\label{Hamiltonian}
H=\hbar\omega_0 \sigma_+\sigma_-+\sum_k \hbar\omega_k b_k^\dag b_k
+\sum_k \hbar (g_k\sigma_+ b_k+g_k^\ast\sigma_- b_k^\dag),
\end{equation}
where $\omega_0$ is the qubit transition frequency, $\sigma_ \pm$ are the qubit raising and lowering operators, the index $k$ labels the environment field modes with frequencies $\omega_k$, $b_k^\dag $ and $b_k $ are the creation and annihilation operators, respectively, and $g_k$ the coupling constants. In the following, we shall give general expressions for the concurrence valid for any local qubit-environment interaction described by the Hamiltonian of Eq.~(\ref{Hamiltonian}). When the environment is at zero-temperature and the qubit is initially in a general superposition state of its two levels, the single-qubit reduced density matrix $\hat{\rho}^S(t)$ ($S=A,B$) takes the form\cite{petru}
\begin{equation}\label{rhoS}
\hat{\rho}^S(t)=\left(%
\begin{array}{cc}
\rho^S_{11}(0)|q(t)|^2  & \rho^S_{10}(0)q(t)\\\\
\rho^S_{01}(0)q^*(t)  & \rho^S_{00}(0)+ \rho^S_{11}(0)(1-|q(t)|^2) \\
\end{array}\right).
\end{equation}
The single-qubit dynamics thus depends only on the function $q(t)$ ($0<|q(t)|<1$, $q(0)=1$), whose explicit time-dependence contains the information on the environment spectral density and the coupling constants. Using Eq.~(\ref{totalevo}), one can then determine the two-qubit dynamics. In particular, in the standard basis $\mathcal{B}=\{\ket{1}\equiv\ket{11},\ket{2}\equiv\ket{10}, \ket{3}\equiv\ket{01}, \ket{4}\equiv\ket{00}\}$, the diagonal elements of the reduced density matrix $\hat{\rho}(t)$ at time $t$ result to be
\begin{eqnarray}\label{rototdiag}
\rho_{11}(t)&=&\rho_{11}(0)|q(t)|^4,\quad
\rho_{22}(t)=\rho_{11}(0)|q(t)|^2(1-|q(t)|^2)+\rho_{22}(0)|q(t)|^2,\nonumber\\
\rho_{33}(t)&=&\rho_{11}(0)|q(t)|^2(1-|q(t)|^2)+\rho_{33}(0)|q(t)|^2,\nonumber\\
\rho_{44}(t)&=&1-[\rho_{11}(t)+\rho_{22}(t)+\rho_{33}(t)],
\end{eqnarray}
and the non-diagonal elements
\begin{eqnarray}\label{rototnodiag}
\rho_{12}(t)&=&\rho_{12}(0)q(t)|q(t)|^2,\ \rho_{13}(t)=\rho_{13}(0)q(t)|q(t)|^2,\ 
\rho_{14}(t)=\rho_{14}(0)q(t)^2,\nonumber\\ 
\rho_{23}(t)&=&\rho_{23}(0)|q(t)|^2,\
\rho_{24}(t)=\rho_{13}(0)q(t)(1-|q(t)|^2)+\rho_{24}(0)q(t),\nonumber\\
\rho_{34}(t)&=&\rho_{12}(0)q(t)(1-|q(t)|^2)+\rho_{34}(0)q(t),
\end{eqnarray}
with $\rho_{ij}(t)=\rho^*_{ji}(t)$, $\hat{\rho}(t)$ being a hermitian matrix. Eqs.~(\ref{rototdiag}) and (\ref{rototnodiag}) give the two-qubit density matrix evolution for any initial state dependent only on $q(t)$.

This way, for the case of the above spin-boson Hamiltonian the concurrence, or any other measure of correlations calculated on the system state, can be shown to depend only on the initial state conditions and on the single-qubit parameter $q(t)$, whose explicit form is determined by the specific spectral density of the environment. This explicit dependence of the concurrence on $q(t)$ has been already reported in the case of initial Bell-like states of Eq.~(\ref{Bell-likestates}), when the system state remains of this form during the evolution (as happens in the case here considered). In particular, these expressions are
\begin{equation}\label{concBell-likestates}
C_\Phi(t)=2\mathrm{max}\{0,a|b||q(t)|^2\},\
C_\Psi(t)=2\mathrm{max}\{0,|b||q(t)|^2[a-|b|(1-|q(t)|^2)]\}.
\end{equation}
The time-dependent forms of concurrences above show that, if the two-qubit system starts from the Bell-like state $\ket{\Phi}$, ESD can never occur being $C_\Phi(t)$ proportional to a power of the single-qubit coherence $|q(t)|$. For initial Bell (maximally entangled) states, $\ket{\Phi^\pm}$ and $\ket{\Psi^\pm}$, the concurrences at the time $t$ take the simple form $C_{\Phi^\pm}(t)=|q(t)|^2$, $C_{\Psi^\pm}(t)=|q(t)|^4$, from which it is readily seen that for both these initial states ESD never occurs. Now we shall extend the expressions of concurrence in terms of $q(t)$ also for the rather more general class of initial EWL states of Eq.~(\ref{EWL}), whose Bell-like states are a subclass. 

During this dynamics, the X-structure of the initial density matrix is maintained and this is the case also for EWL states. The explicit expressions of concurrences at time $t$ are given for the two initial states $\hat{\rho}^\Phi$ and $\hat{\rho}^\Psi$ of Eq.~(\ref{EWL}), respectively, by
\begin{eqnarray}\label{concEWLstates}
C_\rho^\Phi(t)&=&2\mathrm{max}\left\{0,|q(t)|^2\left[ra|b|-\frac{1}{2}\sqrt{(1-r)\left(1-|q(t)|^2+|q(t)|^4\frac{1-r}{4}\right)}\right]\right\},\nonumber\\
C_\rho^\Psi(t)&=&2\mathrm{max}\left\{0,|q(t)|^2\left[ra|b|-\frac{1}{4}(|q(t)|^2(1+3r-4ra^2)+4ra^2-2(1+r))\right]\right\},\nonumber\\
\end{eqnarray}
where $|b|=\sqrt{1-a^2}$. These equations are quite general since their form does not depend on the particular choice of the environment, but only on the Hamiltonian model of Eq.~(\ref{Hamiltonian}) and on the chosen initial state. Once the environment structure is specified, the explicit form of $q(t)$
is obtained that in turns determines the explicit time-dependence of concurrence.

\subsubsection{Bosonic environment with Lorentzian spectral density}
The Hamiltonian of Eq.~(\ref{Hamiltonian}), which may describe various spin-boson systems, is now taken to represent 
a qubit formed by the excited and ground electronic state of a two-level atom interacting with the quantized modes of a high-$Q$ cavity. This dissipative qubit-environment interaction gives the so-called amplitude damping channel\cite{nielsenchuang}.
At zero temperature, this open quantum system has an exact solution\cite{garraway1997}.
In the case of a single excitation in the atom-cavity system, the effective spectral
density $J(\omega)$ is taken as the spectral distribution of an
electromagnetic field inside an imperfect cavity supporting the mode
$\omega_0$, resulting from the combination of the reservoir spectrum
and the system-reservoir coupling, with $\Gamma$ related to the
microscopic system-reservoir coupling constant. This spectral density has the Lorentzian form\cite{petru}
\begin{equation}\label{spectraldensity}
J(\omega)=(\Gamma\lambda^2/2 \pi)/[(\omega_0-\omega)^2+\lambda^2],
\end{equation}
where $\lambda$, defining the spectral width of the coupling, is
connected to the reservoir correlation time $\tau_B$ by the relation
$\tau_B \approx \lambda^{-1}$. In fact, it can be shown that the
reservoir correlation function, corresponding to the spectral
density of Eq.~(\ref{spectraldensity}), has an exponential form with
$\lambda$ as decay rate. On the other hand the parameter $\Gamma$
can be shown to be related to the decay of the excited state of the
atom in the Markovian limit of flat spectrum. The relaxation time
scale $\tau_R$ over which the state of the system changes is then
related to $\Gamma$ by $\tau_R \approx \Gamma^{-1}$. The exact
solution for the single-qubit density matrix (at zero temperature)
with the spectral density given by Eq.~(\ref{spectraldensity}) has
the form of Eq.~(\ref{rhoS}) with\cite{petru,maniscalco}
\begin{equation}\label{qnonmark}
|q(t)|=\mathrm{e}^{-\lambda t/2}\left[\cos (dt/2)+(\lambda/ d)\sin (dt/2)\right],
\end{equation}
where $d=\sqrt{2\Gamma \lambda-\lambda^2}$. A weak and a strong
coupling regime can then be distinguished for the single qubit dynamics.
The weak regime corresponds to the case $\lambda/\Gamma>2$, that is
$\tau_R>2\tau_B$. In this regime the relaxation time is greater than
the reservoir correlation time and the dynamics is essentially
Markovian with an exponential decay controlled by $\Gamma$. In the
strong coupling regime, that is for $\lambda/\Gamma\ll2$, or $\tau_R\ll
2\tau_B$, the reservoir correlation time is greater or of the same
order of the relaxation time and non-Markovian effects become
relevant. Substituting the explicit expression
of $|q(t)|$ of Eq.~(\ref{qnonmark}) in
Eqs.~(\ref{rototdiag}) and (\ref{rototnodiag}), the
two-qubit density matrix at time $t$ and its concurrence are obtained for both the initial states of
Eq.~(\ref{EWL}). 

\begin{figure}[bt]
\centerline{\psfig{file=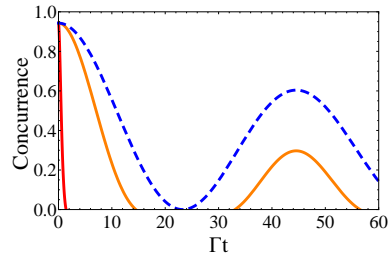,width=5 cm}}
\vspace*{8pt}
\caption{\label{fig:concindenv}Concurrence as a function of $\Gamma t$ for initial Bell-like states $\ket{\Phi}$ (blue dashed line) and $\ket{\Psi}$ (orange solid line) for the case of independent non-Markovian environments with $\lambda/\Gamma=0.01$. The red solid curve on the left represents the concurrence for $\ket{\Psi}$ under Markovian condition $\lambda/\Gamma=3$. The initial Bell-like states have $a^2=1/3$,}
\end{figure}

The time behavior of the concurrences $C_{\Phi}(t)$ and $C_{\Psi}(t)$ as
a function of the dimensionless time $\Gamma
t$ are plotted in Fig.~\ref{fig:concindenv} .
For Markovian conditions ($\lambda/\Gamma>2$) the phenomenon of ESD is retrieved\cite{yu2009Science}, that is entanglement disappears at a finite time, as also observed both in an all-optical experiment\cite{almeida} and in two atomic ensembles\cite{kimble2007PRL}. When non-Markovian effects are relevant ($\lambda/\Gamma\ll 2$), revivals of entanglement may appear after finite periods of time
when disentanglement is complete. This condition of non-Markovianity can be achieved, for example, in cavity QED experimental configurations by using Rydberg atoms with lifetimes $T_\textrm{at}\approx30\textrm{ms}$, inside
Fabry-Perot cavities with quality factors $Q\approx
4.2\times10^{10}$ corresponding to cavity lifetimes\cite{kuhr}
$T_\textrm{cav} \approx 130\textrm{ms}$: these values indeed
correspond to $2\lambda/\Gamma \approx 0.2$. The phenomenon of revivals is induced by the memory of the reservoir, which allows the two-qubit
entanglement to reappear after a dark period of time, during which
the concurrence is zero. These revivals, although they effectively extend the possible usage time of entanglement, decrease with time and eventually disappear after a certain critical time. 
In conclusion, revivals of entanglement, after finite periods of
entanglement disappearance, are linked to the
qubit-environment non-Markovian dynamics and, after their discovery, they have been explained in terms of transfer of correlations back and forth from the two-qubit system to the various parts of the total system. This correlation transfer can be seen as due to the back-action via the environment on the system, which creates correlations between qubits and environments and between the environments themselves: in particular, correlations may build up between the two independent quantum reservoirs\cite{lopez2008PRL,bai2009PRA,Lopez2010PRA} (this phenomenon has been named sudden birth of entanglement in reservoirs). In this sense, the nature of the entanglement revivals described above, occurring for bosonic independent environments\cite{bellomo2007PRL,bellomo2008PRA}, differs from that of revivals obtained for interacting qubits embedded in a common Markovian bosonic reservoir\cite{horodecki2001PRA,tanas}, where revivals are just due to a direct dipole-dipole interaction among. Collapse and revival features of the entanglement dynamics of different
polarization-entangled photon states have been observed in an all-optical experimental setup, when one of the travelling photons passes through a non-Markovian quantum environment simulated by a suitable Fabry-Perot cavity followed by quartz plates\cite{Xu2010PRL}. It has been also recently seen that only ESD with no revivals may occur in the case of independent non-Markovian baths with broadband spectra typical of solid-state nanodevices\cite{palermocatania2010PRA,palermocatania2011IJQI}. Recently it has been shown that the presence of the system-bath coherence in the initial state may change dramatically the entanglement dynamics of two qubits in the presence of a non-Markovian quantum-mechanical bath\cite{Dijkstra2010PRL}. Entanglement dynamics of a system of N qubits interacting with N independent reservoirs in both Markovian and non-Markovian regimes has been also investigated. Entanglement revivals of qubits at instantaneous points of
disappearance or after a finite interval of abrupt vanishing due to the memory effect of non-Markovian reservoirs have been pointed out\cite{Man2010NJP}.

Also the dynamics of nonlocal correlations, identified by violations of the CHSH-Bell inequality\cite{horodecki1995PLA,bellomo2010PLA}, has been studied in independent non-Markovian bosonic environments, like the ones described here, and compared with the dynamics of entanglement\cite{bellomo2008bell}. Revivals of Bell inequality violations have been thus found but they show up for quite strong non-Markovian features. In particular, there may exist time regions when the Bell inequality is not violated even in correspondence with high values of concurrence ($C\sim0.8$: this indicates that even highly entangled systems cannot be exploited with certainty in contexts where the presence of nonlocal correlations is required (for instance, in secure quantum cryptography\cite{acin2006PRL,gisin2007natphoton}).

An all-optical experimental setup has been recently employed to observe and engineer entanglement oscillations of a pair of polarization qubits in an effective non-Markovian channel. Programmable oscillations of entanglement are achieved, where the entangled state obtained at the maximum of the revival after sudden death violates Bell inequality by 17 standard deviations\cite{Paris2011PRA}.

\subsection{Common bosonic environment}
We now review the main features of entanglement dynamics when two noninteracting qubits are embedded in a common non-Markovian bosonic environment (see right part of Fig. 1). As the literature of this topic is extensive, we will focus here on a simple model of dissipative dynamics\cite{maniscalco2008PRL,mazzola2009PRA}.

The Hamiltonian for the total system, in the dipole and
the rotating-wave approximations, and in units of $\hbar = 1$, can be written as 
\begin{eqnarray}
H = \omega_1 \sigma^{(1)}_+  \sigma^{(1)}_- + \omega_2
\sigma^{(2)}_+\sigma^{(2)}_- +  \sum_k \omega_k b_k^{\dag} b_k
+  \left( \alpha_1 \sigma^{(1)}_+   +
\alpha_2\sigma^{(2)}_+ \right) \sum_k g_k b_k  + {\rm h.c.},
\label{eq:Hint}
\end{eqnarray}
where $b^\dag_k$, $b_k$ are the creation and annihilation
operators of quanta of the reservoir,  $ \sigma^{(j)}_{\pm}$ and
$\omega_{j}$ are the inversion operators and the transition frequency
of the $j$-th qubit  ($j=1,2$), respectively, $\omega_k$ is the frequency of
the reservoir $k$-th mode, and
 $\alpha_j g_k$ describe the coupling strength between the
$j$-th qubit and the $k$-th mode of the reservoir.

Here, $\alpha_j$ are dimensionless real coupling constants
measuring the interaction strength of each single qubit with the
reservoir. In the case of two atoms inside a cavity, e.g., different values of $\alpha_j$ can be obtained by changing the relative position of the atoms in the
cavity field standing wave. We denote with
$\alpha_T=(\alpha_1^2+\alpha_2^2)^{1/2}$ the collective coupling
constant and with $r_j=\alpha_j/\alpha_T$ the relative interaction
strength.


\subsubsection{One-excitation dynamics}

We begin by considering the case in which only one excitation is
present in the system and the reservoir is in the vacuum. We consider the initial state
\begin{equation}
\ket{\Psi(0)} = \Bigl [ c_{01} \ket{1}_1\ket{0}_2 + c_{02}
\ket{0}_1\ket{1}_2\Bigr] \bigotimes_k
\ket{0_k}_R,\label{initialstate}
\end{equation}
where $c_{01}$ and $c_{02}$ are complex numbers,  $\ket{0}_j$ and $\ket{1}_j$
$(j=1,2)$ are the ground and excited state of the $j$-th qubit,
respectively, and $\ket{0_k}_R$ is the state of the reservoir with
zero excitations in the $k$-th mode.

The time evolution of the
total system is given by
\begin{eqnarray}
\vert \Psi (t) \rangle =  c_1 (t) \vert 1 \rangle_1 \vert 0
\rangle_2 \vert 0 \rangle_R + c_2(t)  \vert 0 \rangle_1 \vert 1
\rangle_2 \vert 0 \rangle_R+  \sum_k c_k (t) \vert
0 \rangle_1 \vert 0 \rangle_2 \vert 1_k \rangle_R, \label{eq:psi}
\end{eqnarray}
$\vert 1_k \rangle_R$ being the state of the reservoir with only
one excitation in the $k$-th mode.

In the standard basis, the reduced density matrix, obtained from the
density operator $\vert \Psi(t) \rangle \langle \Psi(t) \vert$ after
tracing over the reservoir degrees of freedom, takes the form
\begin{equation}
\label{eq:rhos} \rho(t) = \left(
  \begin{array}{cccc}
    0& 0 & 0 & 0 \\
    0&|c_1(t)|^2&c_1(t)c_2^*(t)& 0\\
    0&c_1^*(t)c_2(t)&|c_2(t)|^2& 0\\
    0&0&0&1-|c_1|^2-|c_2|^2\\
  \end{array}
\right).
\end{equation}
The two-qubit dynamics is therefore completely characterized by the
amplitudes $c_{1,2}(t)$. For certain specific structures of the reservoir
one can obtain the exact analytical expressions of $c_{1,2}(t)$ by the Laplace
transform method. 

We consider a structured reservoir describing the electromagnetic field inside a lossy cavity. This
case can be modeled by a Lorentzian broadening of the fundamental
mode cavity, as given by Eq.~(\ref{spectraldensity}). In Ref.\cite{maniscalco2008PRL} the exact, and therefore non-Markovian, analytical
expression for the amplitudes $c_{1,2}(t)$ has been derived. Using these quantities, one can readily calculate the time evolution of entanglement, as measured by concurrence, which for our system is simply given by $C(t) = 2 \left| c_1(t) c_2^*(t) \right|$. An interesting feature of the dynamics is the existence of a non-zero stationary value
of $C$ due to the entanglement of the decoherence-free state $\vert \psi_-\rangle$. The maximum stationary entanglement $C_s^{\rm max} \simeq 0.65$ is obtained for initially
factorized states. 

For weak couplings and/or bad cavity and for an initially separable state, the concurrence increases
monotonically up to its stationary value; whereas, for initially entangled states, the concurrence first goes to zero before
increasing towards $C_s$. The strong coupling/good cavity case  is more rich and presents entanglement oscillations and
revival phenomena for every initial atomic states. One can prove
that for maximally entangled initial states  the revivals
have maximum amplitude when only one of the two atoms is
effectively coupled to the cavity field, i.e. for $r_1=0,1$. This revivals are similar to those present in the independent environments case discussed in Sec. \ref{sec:inden}, as they are also due to the reservoir memory effect. However, in the case of common environment the revivals are enhanced and the amount  of revived entanglement is huge, being comparable to the previous maximum.

Based on this dynamics, a detailed analysis of the experimental implementation of cavity-mediated entanglement generation in the context of trapped ion cavity QED has been presented in Ref.\cite{harkonen2009PRA}. The rich dynamics of the system, showing quantum beats and other truly nonclassical phenomena, has been studied in the off-resonant regime in Ref.\cite{francica2009PRA}. Finally, it has been demonstrated that the loss of entanglement can be completely suppressed by a series of appropriate projective measurements, a consequence of the quantum Zeno effect\cite{maniscalco2008PRL,francica2010PRA}.

\subsubsection{Two-excitation dynamics}

Let us now consider the case in which two excitations are initially present in the system. In this case the derivation of the exact analytical
solution is more complicated. For the sake of simplicity, we assume that two identical atoms interact resonantly with a Lorentzian structured reservoir with equal coupling. In Ref.\cite{mazzola2009PRA} it has been shown that, by using the pseudomode approach\cite{garraway1997} and establishing a
connection with a three-level ladder system, it is still possible to solve the dynamics without performing any approximations. Armed with the exact solution we can prove that ESD and entanglement sudden birth (ESB) phenomena occur in the non-Markovian common reservoir scenario\cite{mazzola2009PRA,mazzola2010JPB}.

Entanglement sudden death and entanglement revivals are basically due to two combined and
intertwined effects: the backaction of the structured reservoir and
the reservoir-mediated interaction between the qubits. In order
to understand the role played by each of these effects it is useful to
compare the dynamics with the cases of the common Markovian
reservoir, and two independent non-Markovian reservoirs.

In a common Markovian reservoir a period of complete disentanglement is followed
by a revival of entanglement, but no oscillations are present. Such
a revival is due to the action of the common reservoir which tends
to create quantum correlations between the qubits, providing an
effective coupling between them. The feedback of information from the reservoir into the system,
characterizing the non-Markovian dynamics, enhances the appearance of
ESD regions, since it tends to recreate the conditions that led to
the first ESD period.

A comparison between the non-Markovian independent and common reservoir cases reveals that, for
the same type of reservoir spectrum, ESD (dark periods) regions are much wider in
the independent reservoirs case than in the common reservoir case.
Since both cases take into account memory effects, this suggests
that the reservoir-mediated interaction between the qubits, in the
common reservoir scenario, effectively counters the fast
disappearance of entanglement.

Non-Markovian effects also influence strongly the dynamics for
initially factorized states, when the reservoir-mediated interaction
between the qubits leads to entanglement generation. While, for independent reservoirs, sudden birth of
entanglement does not appear, for the common reservoir case,
ESB does occur.
Moreover, in a structured reservoir such a phenomenon presents new
interesting features, compared to the Markovian case. In general, the reservoir memory prolongs
the initial disentanglement. In this case, therefore, the reservoir
backaction dominates over the tendency of the common reservoir to
create entanglement between the qubits. ESB periods and
disentanglement revivals become more frequent and numerous for
stronger non-Markovian conditions.

\section{Entanglement trapping in independent structured environments}
Strategies for preventing the decay of quantum correlations, initially present in a composite quantum system, and thus to fight decoherence are important for effective realization of quantum computers. Some of these strategies, applied to a single qubit, are quantum error correction\cite{nielsenchuang}, dynamical decoupling of the quantum system from its environment\cite{viola1998PRA,vandersypen2005RMP} or quantum Zeno effect\cite{facchi2001PRL}. All of these procedures require external controls. It has been then shown that entanglement protection is possible by dynamical decoupling based on nonlocal pulses\cite{muhktar2010PRA1} and by Zeno effect on qubits in a common bosonic reservoir\cite{maniscalco2008PRL,francica2010PRA}.
On the other hand, another possible way to reduce the detrimental effect of the environment, without using external controls, could consist in embedding the qubits in suitably structured environments where, for example, population trapping is achievable. This is known to happen in photonic band-gap (PBG) materials or photonic crystals\cite{yablonovitch1987PRL,john1990PRL}. Here we briefly review the results obtained for two noninteracting qubits in independent photonic crystals\cite{bellomo2008trapping,bellomo2010PhysScrManiscalco}.

Each environment is assumed to be a zero-temperature three-dimensional periodic dielectric with isotropic photon dispersion relation $\omega_k$ (photonic crystal)\cite{john1994PRA}. The Hamiltonian model is always of the spin-boson type as in Eq.~(\ref{Hamiltonian}) with a noise spectral density typical of a PBG material. By symmetrizing $\omega_k$, one produces photonic band gaps at the spheres $|\mathbf{k}|=m\pi/L$ $(m=1,2,\ldots)$, where $L$ is the lattice constant. In such ideal photonic crystals, a PBG is the frequency range over which the local density of electromagnetic states and the decay rate of the atomic population of the excited state vanish. Near the band-gap edges the density of states becomes singular\cite{john1990PRL}, the atom-field interaction becomes strong and one can expect modifications to the spontaneous emission decay. For $k\cong \pi/L$ the dispersion relation near the band edge frequency $\omega_c$ can be approximated by $\omega_k=\omega_c+D(k-\bar{k})^2$, where\cite{john1990PRL} $D\cong\omega_c/\bar{k}^2$. This dispersion relation is isotropic since it only depends on the magnitude $k$ of the wave vector $\mathbf{k}$. In the case of a two-level atom, with atomic transition frequency $\omega_0$, embedded in such a material, the explicit form of single-qubit coherence $q(t)=f(\delta,\beta,t)$ has been obtained\cite{john1994PRA} in terms of the detuning $\delta=\omega_0-\omega_c$ and of a parameter $\beta$ defined as $\beta^{3/2}=\omega_0^{7/2}d^2/6\pi\epsilon_0\hbar c^3$, where $\epsilon_0$ is the Coulomb constant and $d$ the atomic dipole moment. The parameter $\beta^{3/2}$, being proportional to $\omega_0^2d^2$, represents the strength of the atom-reservoir coupling. The time evolution is scaled by $\beta$ and for large times ($\beta t\gg1$) one has $|q(t)|^2\rightarrow\textrm{constant}$, that is a population trapping\cite{john1994PRA}.

The dynamics of entanglement is obtained by the expressions of the concurrences in terms of $q(t)=f(\delta,\beta,t)$, as for example those of Eq.~(\ref{concBell-likestates}) when the two qubits are prepared in a Bell-like state.
\begin{figure}[bt]
{\centerline{\psfig{file=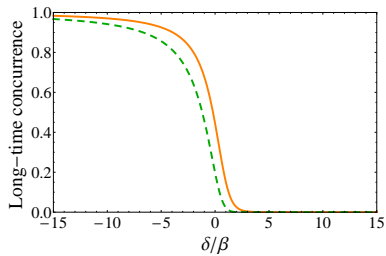,width=5 cm}}}
\vspace*{8pt}
\caption{\label{fig:ConcPBG} Asymptotic values of concurrences $C_\Phi(t=\infty)$ (orange solid line) and $C_\Psi(t=\infty)$ (green dashed line) as a function of $\delta/\beta $, starting respectively from the initial Bell state $\ket{\Phi^\pm}$ and $\ket{\Psi^\pm}$.}
\end{figure}
The concurrence is nearer to its maximum value 1 for atomic frequencies farther from the band edge and deeper inside the gap. In Fig.~\ref{fig:ConcPBG} the asymptotic values of $C_{\Phi^\pm}(t\rightarrow\infty)$ and $C_{\Psi^\pm}(t\rightarrow\infty)$ are plotted as a function of the ratio $\delta/\beta $ and it can be appreciated that $C_{\Phi^\pm}(t\rightarrow\infty)$ is slightly larger than $C_{\Psi^\pm}(t\rightarrow\infty)$. We see that, when $\delta/\beta<0$, these asymptotic values differ from zero and decrease very rapidly going near the edge. It is evident that the more the atomic transition frequency is far from the band edge and inside the gap ($\delta/\beta<0$), the higher is the preserved entanglement. An isotropic dispersion relation is known to lead also to the formation of atom-photon bound states\cite{john1990PRL,john1994PRA}, that in turn lead to entanglement trapping near the edge even for $\omega_0$ outside the band gap ($\delta/\beta>0$), although with small asymptotic values ($<0.4$; see Fig.~\ref{fig:ConcPBG}). For an initial Bell-like state one finds that smaller initial entanglement leads to lower asymptotic value of concurrence and that the entanglement trapping phenomenon does not depend on the initial value of $a$. It is worth to point out that, in a zero-temperature Markovian environment, the initial Bell-like state $\ket{\Psi}$ of Eq.~(\ref{Bell-likestates}) would undergo entanglement sudden death\cite{yu1} for $a<1/\sqrt{2}$; differently, in a PBG material entanglement sudden death and its consequent limit on the entanglement usage can be prevented\cite{bellomo2008trapping}.

The results described above are valid for ideal PBG materials, thus one may ask to what extent they hold for real crystals. In fact, in real crystals with finite dimensions a pseudogap is typically obtained with a density of states, although not exactly zero, strongly smaller than that of free space. For quantum dots inside a real PBG material, for example, a pronounced inhibition of spontaneous emission up to $\sim30\%$ has been experimentally observed\cite{lodhal2004Nature} which would give, on the basis of the relation between population and concurrence given by Eqs.~(\ref{concEWLstates}) and (\ref{concBell-likestates}), a partial inhibition of entanglement decay.

Other analyses have been done, after the above one\cite{bellomo2008trapping}, for two-qubit entanglement in one-dimensional photonic crystals\cite{gaoxiang2009PRA}, in a common PBG material and with a dipole-dipole interaction\cite{zhang2010EPJD,wang2011OptComm}, where sudden death and sudden birth of entanglement have been found. 
We also point out that nonlocality, identified by Bell inequality violations, can be also efficiently protected in photonic crystals\cite{bellomo2009ASL}. The dynamics of quantum discord has been also recently studied in a common PBG material, finding that the quantum discord can maintain a constant value in the long-time limit even when entanglement suddenly disappears\cite{wang2012OptComm}.

\section{Dynamics of correlations other than entanglement}
Here we consider the dynamics of quantum correlations, identified by quantum discord, under different conditions. Relations among quantifiers of entanglement, nonlocality and purity have been studied in dynamical contexts\cite{mazzolapalermo2010PRA} to better understand if and how these quantum properties can be connected to each other. Moreover, it has been shown that both quantum and classical correlations can be protected via the quantum Zeno effect\cite{francica2010PRA}.

\subsection{Frozen discord in non-Markovian dephasing channels}
In the last years a huge deal of attention has been devoted to the dynamics of correlations other than entanglement in presence of both Markovian\cite{werlang2009PRA} and non-Markovian\cite{fanchini2010PRA,wang2010PRA,bellomo2011IJQI,mazzola2011frozen} quantum environments. For a comprehensive review on quantum discord we refer to Ref.\cite{Modi2011arxiv}. In general quantum discord is more robust than entanglement against decoherence, and it does not present sudden death phenomena. 

A peculiar aspect of the dynamics of quantum discord is that it may remain frozen for finite time intervals\cite{mazzola2010PRL,mazzola2011frozen} even in presence of decoherence, as also proven in an all-optical experimental setup\cite{can-guo2010NatComm,xu2010PRA}. In fact, it has been very recently theoretically demonstrated that, for non-Markovian dephasing, quantum discord can be eternally frozen or time invariant\cite{haikka2012arxiv}. This new phenomenon is indeed quite unique, as discord is the only nonclassical property which may remain constant in time in presence of decoherence even when the state of the system evolves due to the interaction with the environment. If, as several studies strongly indicates, quantum discord is a resource for certain quantum technologies, its time invariant properties might play an important role.

The physical system under investigation consists of two qubits under local dephasing channels.
Let us assume that the qubits, $A$ and $B$, are prepared in a state of the form 
\begin{equation}\label{marginals}
\rho_{AB}=\frac{1}{4} \left( \mathbf{ 1}_{AB} + \sum_{i=1}^3 c_i \sigma_{i}^{A}\sigma_{i}^{B}\right), 
\end{equation}
where $\sigma_{j}^{A(B)}$ is the Pauli operator in direction $j$ acting on $A (B)$, $c_{i}$ is a real number, such that
$0\leq|c_{i}|\leq1$ for every $i$, and $\mathbf{ 1}_{AB}$ is the
identity operator of the total system. For a dephasing channel subjected to either white or colored noise the form of the state is maintained during the evolution. 
In the case of white noise the Markovian dissipator takes the form ${\cal L}[\rho_{A(B)}]=\gamma [\sigma^{A(B)}_j \rho_{A(B)} \sigma_j^{A(B)} - \rho_{A(B)}]/2$, with $\gamma$ the phase damping rate, and $j=1,2,3$ for the bit, bit-phase, and phase flip cases, respectively. For certain classes of states the discord does not decay at all, until a finite transition time $\bar{t}$. The system indeed experiences two different dynamical regimes: at the beginning the state of the qubits evolves in a way to lose only classical correlations while quantum correlations are not touched in any way by decoherence. This is what we call the classical decoherence regime. In the second phase, after a sudden change in quantum and classical correlations, quantum correlations are lost while classical correlations are preserved. This is the quantum decoherence phase. This effect was named sudden transition from classical to quantum decoherence.

When the two quits are subjected to local colored noise dephasing channels one needs to use a memory-kernel master equation to study the dynamics of each qubit
subjected to random telegraphic noise\cite{mazzola2011frozen}. In this case, for certain states and regimes of parameters, one can distinguish two
different behaviours for the dynamics of the quantum correlation: the frozen behaviour, and the sudden change dynamics. In the first case the dynamics
displays multiple sudden transitions: as in the Markovian case, classical correlation at first decreases while the discord remains constant then at a sudden transition point classical correlation becomes constant and discord oscillates, and then again discord becomes constant while classical correlation oscillates. In the second case first both classical and quantum correlations decrease until, at a
sudden change point, classical correlations become constant while quantum
correlations exhibit a discontinuous change in the amplitude of the damped
oscillations. At the following sudden change point classical correlation starts again to oscillate and
quantum discord changes back to its previous rate of oscillation. This behavior reflects the presence of memory effects characterizing the dynamics.

\subsection{Revivals of quantum correlations without back-action}
Here we treat the case when environments are classical, where back-action is absent. One would expect that if the quantum environment of a quantum system is modelled as classical, for example a classical light field coupled to a quantized atom, then there might be qualities of the dynamics that this model is unable to describe. Intuitively, one might expect that since a classical environment should not be able to store quantum correlations in the same way as a quantum environment, the capability to describe revivals of quantum correlations might be affected. Similarly, if the environment has no back-action on the system, then it would seem that this could also affect the correlation dynamics. Nevertheless, revivals of entanglement can occur for random classical telegraph noise\cite{zhou2010QIP,lofranco2012PhysScripta}, for atoms subject either to local phase-noisy lasers\cite{bellomo2012PhysScripErika} or to random external classical fields\cite{lofranco2012PRA}. It is clearly important to exactly know how correlation dynamics behaves when an environment is modelled as classical instead of quantum, or when back-action on the system is not present. We now briefly review a recent result\cite{lofranco2012PRA} that tries to explain how and why revivals of correlations, including quantum correlations, can generally occur also if the environment is classical, and when no back-action is present. 
We illustrate our discussion with a model characterized by the absence of correlations induced by back-action, where the element of randomness is introduced in a very simple way. 

The example we consider is a pair of independent qubits driven by single classical field modes with random phase, where it has been shown that revivals of correlations do occur both for entanglement, for the so-called quantum discord, and for classical correlations\cite{lofranco2012PRA}. The dynamical map for the single qubit $S=A,B$ is of the random external fields type\cite{alickibook,horodecki2001PRA} and can be written as
$\Lambda_S(t,0)\rho_S(0)=\frac{1}{2}\sum_{i=1}^2U_i^{S}(t)\rho_S(0)U_i^{S\dag}(t)$,
where $U_i^{S}(t)=\mathrm{e}^{-\mathrm{i}H_it/\hbar}$ is the time evolution operator with $H_i=\mathrm{i}\hbar g(\sigma_+e^{-\mathrm{i}\phi_i}-\sigma_-e^{\mathrm{i}\phi_i})$ and the factor $1/2$ arises from the equal field phase probabilities of the model (more in general, there is a probability $p_i^S$ associated to $U_i^S$). Each Hamiltonian $H_i$ is expressed in the rotating frame at the qubit-field resonant frequency $\omega$. In the basis $\{\ket{1},\ket{0}\}$, the time evolution operators $U_i^{S}(t)$ have the matrix form
\begin{equation}\label{unitarymatrix}
U_i^{S}(t)=\left(
\begin{array}{cc}\cos(gt)&\mathrm{e}^{-\mathrm{i}\phi_i}\sin(gt)\\
-\mathrm{e}^{\mathrm{i}\phi_i}\sin(gt) & \cos(gt) \\\end{array}\right),
\end{equation}
where $i=1,2$ with $\phi_1=0$ and $\phi_2=\pi$.
\begin{figure}[bt]
{\centerline{\psfig{file=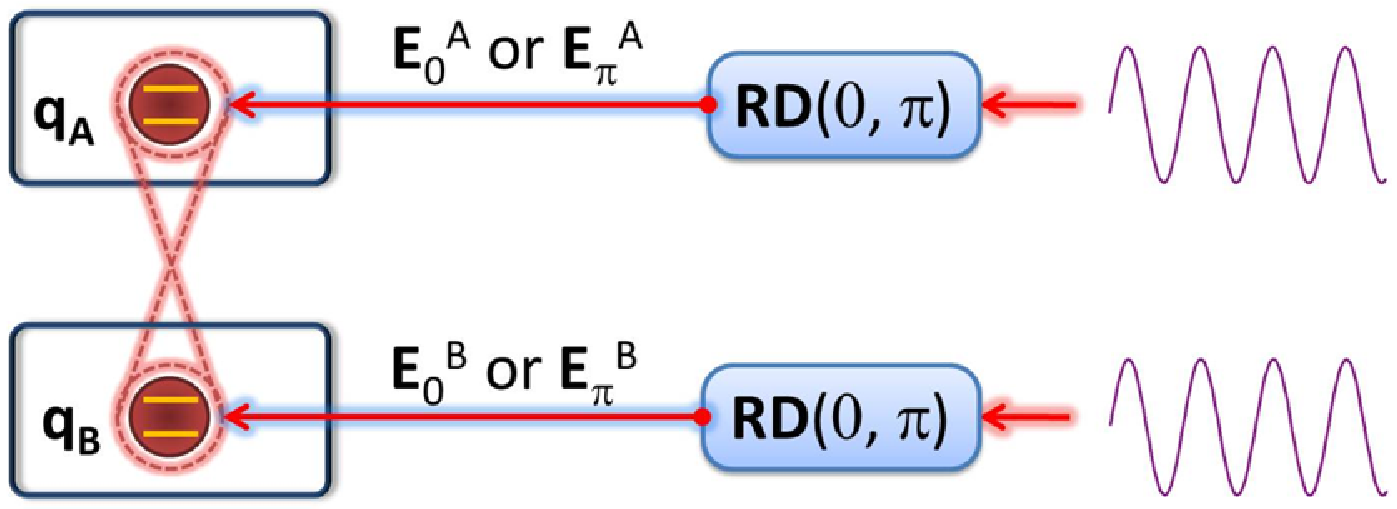,width=6 cm}\hspace{1 cm}
\psfig{file=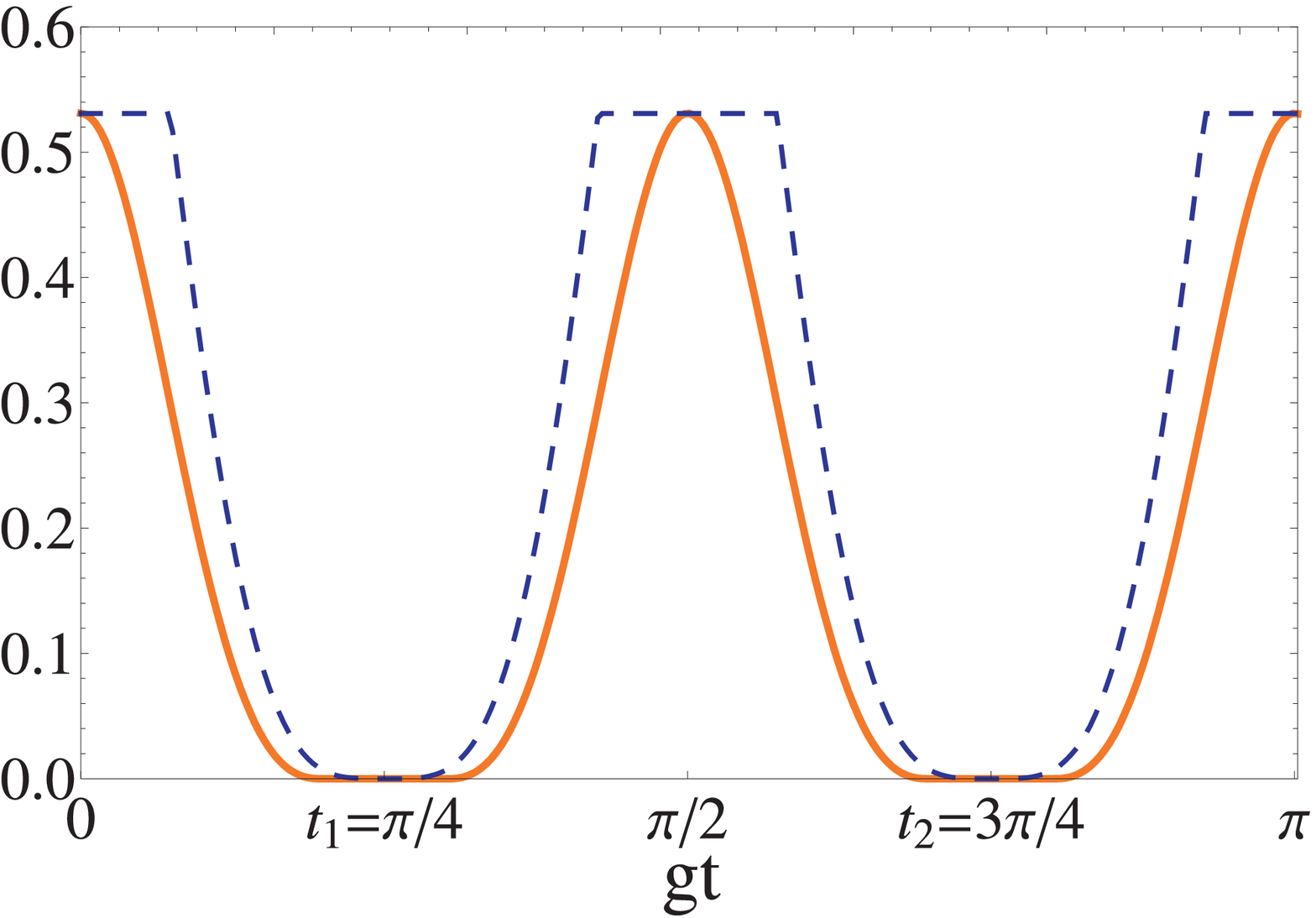,width=5 cm}}}
\vspace*{8pt}
\caption{\label{fig:randomexternalfields} \textbf{Left figure.} Sketch of the physical system. The two qubits are initially entangled. Each random dephaser (RD) is such that the field phase at the location of each qubit is either zero or $\pi$ with equal probability, $p=1/2$, corresponding to either $\mathbf{E}^S_{0}$ or $\mathbf{E}^S_{\pi}$ ($S=A,B$). \textbf{Right figure.} Dynamics of entanglement $E$ (orange solid line) and discord $D$ (blue dashed line) for an initial Bell-diagonal state $\rho_\mathrm{B}(0)=0.9\ket{\Phi^+}\bra{\Phi^+}+0.1\ket{\Phi^-}\bra{\Phi^-}$. $D$ vanishes only at given times $t_n=(2n-1)\pi/(4g)$ ($n=1,2,\ldots$), while $E$ disappears for finite time intervals. Figures reproduced from [R. Lo Franco \textit{et al.}, \textit{Phys. Rev. A} \textbf{85}, 032318 (2012)].}
\end{figure}
This simple model is depicted in the left part of Fig.~\ref{fig:randomexternalfields}, where the two local environments are realized by two single classical field modes of the same frequency and amplitude, but each passing across a random dephaser such that the phase of the field at the location of each qubit is either zero or $\pi$, with probability $p=1/2$. The interaction between each qubit and its local field mode is assumed to be strong enough so that, for sufficiently long times, the dissipation effects of the vacuum radiation modes on the qubit dynamics can be neglected. The dynamics is cyclic, as follows from the unitary matrix of Eq.~(\ref{unitarymatrix}) (for each $i=1,2$) becoming the identity for times $t=k\pi/g$ ($k=0,1,2,\ldots$). The same initial state is thus retrieved at these times. If one takes into account dissipation effects on the system, the dynamics would not be cyclic anymore, but it is expected that for dissipation weak enough the qualitative behavior of the dynamics would remain the same until a characteristic time.

The global dynamical map $\Lambda$ applied to an initial state $\rho(0)$ of the bipartite system, $\rho(t)\equiv\Lambda(t,0)\rho(0)$, is again of the random external fields type, that is,
\begin{equation}\label{globalrandomfieldmap}
\rho(t)=\frac{1}{4}\sum_{i,j=1}^2U_i^{A}(t)U_j^{B}(t)\rho(0)U_i^{A\dag}(t)U_j^{B\dag}(t).
\end{equation}
This map $\Lambda$ moves inside the class of Bell-diagonal states, that are mixture of the four Bell states. Given a state $\rho$, we adopt the definitions\cite{Modi2010PRL} of entanglement $E(\rho)\equiv S(\rho\|\sigma_\rho)$ and quantum discord $D(\rho)\equiv S(\rho\|\chi_\rho)=S(\chi_\rho)-S(\rho)$, where $S(\rho\|\sigma)\equiv-\mathrm{Tr}(\rho\log \sigma)-S(\rho)$ is the relative entropy, $S(\rho)\equiv-\mathrm{Tr}(\rho\log\rho)$ is the von Neumann entropy, $\chi_\rho$ and $\sigma_\rho$ are, respectively, the classical state and the separable state closest to $\rho$.  These quantities can be analytically calculated for Bell-diagonal states. Right plot of Fig.~\ref{fig:randomexternalfields} displays the dynamics of entanglement $E$ and discord $D$ starting from the Bell-diagonal state $\rho_\mathrm{B}(0)=0.9\ket{\Phi^+}\bra{\Phi^+}+0.1\ket{\Phi^-}\bra{\Phi^-}$. Collapses and revivals of entanglement occur, as is known to happen for independent non-Markovian bosonic environments acting as amplitude damping channels\cite{bellomo2007PRL}. Quantum discord also decreases and increases during the dynamics, vanishing only at given times. In presence of small dissipation, the same qualitative behavior of correlations is expected, that is, partial revivals should still occur when the dissipation effects are weak enough. 

The above revivals of quantum correlations, as characterized by entanglement and discord, occur without transfer of correlations induced by back-action from system to environment. The absence of back-action implies that the initial two-qubit correlations cannot be ``temporarily stored" as correlations entirely within the environments. This differs from the point of view where revivals are interpreted in terms of transfer of correlations induced by back-action which correlate the independent non-Markovian quantum environments, which in turn again re-correlate the qubits\cite{bellomo2007PRL,lopez2008PRL,bai2009PRA,Lopez2010PRA}. In our example this mechanism is absent. These revivals can be instead explained in terms of correlations in a classical-quantum state between qubits and environments\cite{lofranco2012PRA}. The environments itself play an important role through the classical record they keep memory of what operation has been applied to the qubits. Finally, it has been also shown that these revivals can be interpreted by referring only to the intrinsic features of the two-qubit dynamics, by showing that they are connected with the increase of a quantifier of non-Markovianity\cite{plenio2010PRL} of the single-qubit dynamics.

\section{Conclusions}

The reduced dynamics of a system interacting with a non-Markovian environment can be characterized by relevant memory effects. In this non-Markovian regime, correlations present in a composite system may evolve in ways which can be exploited in applications requiring their presence. In this paper, we have reviewed some recent results concerning the characterization of different kinds of correlations in non-Markovian environments.

In a first part of this review, we have discussed the phenomenon of entanglement revivals. In the case of two non-interacting qubits initially prepared in an entangled state and subject to local Markovian noise, in front of an exponential decay of single qubit coherences, entanglement may disappear in a finite time. Differently, in the case of non-Markovian environments, it has been shown theoretically and experimentally that entanglement may revive also after finite time intervals when it is absent, because of the memory effects in the two-qubit dynamics.

Following this result, as a possible strategy for preventing the decay of quantum correlations, we have discussed on the possibility to exploit at the maximum non-Markovian effects to protect entanglement from noise. In the case of qubits embedded in photonic crystals, entanglement may be preserved for arbitrary long times, phenomenon due to the direct link between the dynamics of entanglement and that of the excited state population of each qubit; indeed in a photonic-band gap material, spontaneous emission of a qubit may be inhibited and population trapping may occur, as observed for example in the case of quantum dots.

In the final part we have also discussed recent results concerning the occurrence of revival of correlations also in the case of classical environments such that back-action on the system is absent. The mechanism at the basis of these revivals appears to be connected to the memory effects in the reduced dynamics and is linked to a subtle mechanism that the environment itself can play through the classical record they keep memory of what operation has been applied to the qubits. This occurs even if the classical environments cannot store any quantum correlations on their own, and they do not become entangled with their respective quantum systems.

\section*{Acknowledgements}
SM acknowledges financial support from the Finnish Cultural Foundation (Science Workshop on Entanglement), the Emil Aaltonen foundation (Non-Markovian Quantum Information), and EP- SRC (EP/J016349/1).


\section*{References}

\begin{thebibliography}{100}

\bibitem{schrodinger}
E.~Schr{\"o}dinger,
\newblock \emph{Naturwissenschaften} \textbf{23}, 844 (1935)

\bibitem{nielsenchuang}
M.~A. Nielsen, I.~L. Chuang,
\newblock \emph{Quantum Computation and Quantum Information}
\newblock (Cambridge University Press, Cambridge, 2000)

\bibitem{petru}
H.-P. Breuer, F.~Petruccione,
\newblock \emph{The Theory of Open Quantum Systems}
\newblock (Oxford University Press, Oxford, New York, 2002)

\bibitem{benenti}
G.~Benenti, G.~Casati, G.~Strini,
\newblock \emph{Principles of quantum computation and information}
\newblock (World Scientific, Singapore, 2007)

\bibitem{bell}
J.~S. Bell,
\newblock \emph{Physics} \textbf{1}, 195  (1964)

\bibitem{clauser}
J.~F. Clauser, M.~A. Horne, A.~Shimony, R.~A. Holt,
\newblock \emph{Phys. Rev. Lett.} \textbf{23}, 880  (1969)

\bibitem{Zurek2001PRL}
H.~Ollivier, W.~H. Zurek,
\newblock \emph{Phys. Rev. Lett.} \textbf{88}, 017901 (2001) 

\bibitem{vedral2001JPA}
L.~Henderson, V.~Vedral,
\newblock \emph{J. Phys. A} \textbf{34}, 6899 (2001) 

\bibitem{dakic2010PRL}
B.~Daki{\'{c}}, V.~Vedral, {\v{C}}.~Brukner,
\newblock \emph{Phys. Rev. Lett.} \textbf{105}, 190502 (2010) 

\bibitem{bennett}
C.~H. Bennett, G.~Brassard, C.~Crepeau, R.~Jozsa, A.~Peres, W.~K. Wootters,
\newblock \emph{Phys. Rev. Lett.} \textbf{70}, 1895 (1993) 

\bibitem{acin2006PRL}
A.~Acin, N.~Gisin, L.~Masanes,
\newblock \emph{Phys. Rev. Lett.} \textbf{97}, 120405 (2006) 

\bibitem{gisin2007natphoton}
N.~Gisin, R.~Thew,
\newblock \emph{Nature Photon.} \textbf{1}, 165 (2007) 

\bibitem{Knill1998PRL}
E.~Knill, R.~Laflamme,
\newblock \emph{Phys. Rev. Lett.} \textbf{81}, 5672 (1998) 

\bibitem{Lanyon2008PRL}
B.~P. Lanyon, M.~Barbieri, M.~P. Almeida, A.~G. White,
\newblock \emph{Phys. Rev. Lett.} \textbf{101}, 200501 (2008) 

\bibitem{Datta2008PRL}
A.~Datta, A.~Shaji, C.~Caves,
\newblock \emph{Phys. Rev. Lett.} \textbf{100}, 050502 (2008) 

\bibitem{DakicZeilinger2012arXiv}
B.~Daki{\'{c}} \emph{et~al.},
\newblock arXiv:1203.1629

\bibitem{adesso2012arxiv}
T.~Tufarelli \emph{et~al.},
\newblock arXiv:1205.0251

\bibitem{Modi2010PRL}
K.~Modi, T.~Paterek, W.~Son, V.~Vedral, M.~Williamson,
\newblock \emph{Phys. Rev. Lett.} \textbf{104}, 080501 (2010) 

\bibitem{luo2011PRL}
S.~Luo, S.~Fu,
\newblock \emph{Phys. Rev. Lett.} \textbf{106}, 120401 (2011) 

\bibitem{bellomo2012PRA}
B.~Bellomo, G.~L. Giorgi, F.~Galve, R.~{Lo Franco}, G.~Compagno, R.~Zambrini,
\newblock \emph{Phys. Rev. A} \textbf{85}, 032104 (2012) 

\bibitem{bellomo2012linearentropy}
B.~Bellomo, R.~{Lo Franco}, G.~Compagno,
\newblock arXiv:1104.4043

\bibitem{diosi}
L.~Di\'{o}si,
\newblock \emph{Lect. Notes Phys.} \textbf{622}, 157 (2003) 

\bibitem{yu1}
T.~Yu, J.~H. Eberly,
\newblock \emph{Phys. Rev. Lett.} \textbf{93}, 140404 (2004) 

\bibitem{santos}
M.~F. Santos, P.~Milman, L.~Davidovich, N.~Zagury,
\newblock \emph{Phys. Rev. A} \textbf{73}, 040305(R) (2006) 

\bibitem{carv1}
A.~R.~R. Carvalho, F.~Mintert, S.~Palzer, A.~Buchleitner,
\newblock \emph{Eur. Phys. J. D} \textbf{41}, 425 (2007) 

\bibitem{qasimi2008PRA}
A.~Al-Qasimi, D.~F.~V. James,
\newblock \emph{Phys. Rev. A} \textbf{77}, 012117 (2008) 

\bibitem{yu2009Science}
T.~Yu, J.~H. Eberly,
\newblock \emph{Science} \textbf{323}, 598 (2009) 

\bibitem{almeida}
M.~P. Almeida, \emph{et~al.},
\newblock \emph{Science} \textbf{316}, 579 (2007) 

\bibitem{kimble2007PRL}
J.~Laurat, K.~S. Choi, H.~Deng, C.~W. Chou, H.~J. Kimble,
\newblock \emph{Phys. Rev. Lett.} \textbf{99}, 180504 (2007) 

\bibitem{miran2004PLA}
A.~Miranowicz,
\newblock \emph{Phys. Lett. A} \textbf{327}, 272 (2004) 

\bibitem{kofman2008PRA}
A.~G. Kofman, A.~N. Korotkov,
\newblock \emph{Phys. Rev. A} \textbf{77}, 052329 (2008) 

\bibitem{bellomo2007PRL}
B.~Bellomo, R.~{Lo Franco}, G.~Compagno,
\newblock \emph{Phys. Rev. Lett.} \textbf{99}, 160502 (2007) 

\bibitem{bellomo2008PRA}
B.~Bellomo, R.~{Lo Franco}, G.~Compagno,
\newblock \emph{Phys. Rev. A} \textbf{77}, 032342 (2008) 

\bibitem{Xu2010PRL}
J.-S. Xu, \emph{et~al.},
\newblock \emph{Phys. Rev. Lett.} \textbf{1}, 7 (2010) 

\bibitem{bellomo2011PhyScrSavasta}
B.~Bellomo, G.~Compagno, R.~L. Franco, A.~Ridolfo, S.~Savasta,
\newblock \emph{Phys. Scripta} \textbf{T143}, 014004 (2011) 

\bibitem{maniscalco2008PRL}
S.~Maniscalco, F.~Francica, R.~L. Zaffino, N.~{Lo Gullo}, F.~Plastina,
\newblock \emph{Phys. Rev. Lett.} \textbf{100}, 090503 (2008) 

\bibitem{mazzola2009PRA}
L.~Mazzola, S.~Maniscalco, J.~Piilo, K.-A. Suominen, B.~M. Garraway,
\newblock \emph{Phys. Rev. A} \textbf{79}, 042302 (2009) 

\bibitem{bellomo2008bell}
B.~Bellomo, R.~{Lo Franco}, G.~Compagno,
\newblock \emph{Phys. Rev. A} \textbf{78}, 062309 (2008) 

\bibitem{fanchini2010PRA}
F.~F. Fanchini, \emph{et~al.},
\newblock \emph{Phys. Rev. A} \textbf{81}, 052107 (2010) 

\bibitem{wang2010PRA}
B.~Wang, Z.-Y. Xu, Z.-Q. Chen, M.~Feng,
\newblock \emph{Phys. Rev. A} \textbf{81}, 014101 (2010) 

\bibitem{bellomo2011IJQI}
B.~Bellomo, G.~Compagno, R.~L. Franco, A.~Ridolfo, S.~Savasta,
\newblock \emph{Int. J. Quant. Inf.} \textbf{9}, 1665 (2011) 

\bibitem{bellomo2008trapping}
B.~Bellomo, R.~{Lo Franco}, S.~Maniscalco, G.~Compagno,
\newblock \emph{Phys. Rev. A} \textbf{78}, 060302(R) (2008) 

\bibitem{bellomo2009ASL}
B.~Bellomo, R.~{Lo Franco}, G.~Compagno,
\newblock \emph{Adv. Sci. Lett.} \textbf{2}, 459 (2009) 

\bibitem{bellomo2010PhysScrManiscalco}
B.~Bellomo, R.~{Lo Franco}, S.~Maniscalco, G.~Compagno,
\newblock \emph{Phys. Scripta} \textbf{T140}, 014014 (2010) 

\bibitem{yu5}
T.~Yu, J.~H. Eberly,
\newblock \emph{Quantum Information and Computation} \textbf{7}, 459 (2007) 

\bibitem{bose2001}
S.~Bose, I.~Fuentes-Guridi, P.~L. Knight, V.~Vedral,
\newblock \emph{Phys. Rev. Lett.} \textbf{87}, 050401 (2001) 

\bibitem{Pratt2004PRL}
J.~S. Pratt,
\newblock \emph{Phys. Rev. Lett.} \textbf{93}, 237205 (2004) 

\bibitem{peters2004PRL}
N.~A. Peters, J.~B. Altepeter, D.~Branning, E.~R. Jeffrey, T.-C. Wei, P.~G. Kwiat,
\newblock \emph{Phys. Rev. Lett.} \textbf{92}, 133601 (2004) 

\bibitem{wang}
J.~Wang, H.~Batelaan, J.~Podany, A.~F. Starace,
\newblock \emph{J. Phys. B, At. Mol. Opt. Phys.} \textbf{39}, 4343 (2006) 

\bibitem{hagley1997PRL}
E.~Hagley, \emph{et~al.},
\newblock \emph{Phys. Rev. Lett.} \textbf{79}, 1 (1997) 

\bibitem{chiuriPRA2011}
A.~Chiuri, G.~Vallone, M.~Paternostro, P.~Mataloni,
\newblock \emph{Phys. Rev. A} \textbf{84}, 020304(R) (2011) 

\bibitem{dicarlo2009Nature}
L.~DiCarlo, \emph{et~al.},
\newblock \emph{Nature} \textbf{460}, 240 (2009) 

\bibitem{Fazio2002Nature}
A.~Osterloh, L.~Amico, G.~Falci, R.~Fazio,
\newblock \emph{Nature} \textbf{416}, 608 (2002) 

\bibitem{osborne2002PRA}
T.~J. Osborne, M.~A. Nielsen,
\newblock \emph{Phys. Rev. A} \textbf{66}, 032110 (2002) 

\bibitem{munro}
W.~J. Munro, D.~F.~V. James, A.~G. White, P.~G. Kwiat,
\newblock \emph{Phys. Rev. A} \textbf{64},030302(R)  (2001) 

\bibitem{wei}
T.~C. Wei, \emph{et~al.},
\newblock \emph{Phys. Rev. A} \textbf{67}, 022110 (2003) 

\bibitem{wootters}
W.~K. Wootters,
\newblock \emph{Phys. Rev. Lett.} \textbf{80}, 2245 (1998) 

\bibitem{garraway1997}
B.~M. Garraway,
\newblock \emph{Phys. Rev. A} \textbf{55}, 2290 (1997) 

\bibitem{maniscalco}
S.~Maniscalco, F.~Petruccione,
\newblock \emph{Phys. Rev. A} \textbf{73}, 012111 (2006) 

\bibitem{kuhr}
S.~Kuhr, \emph{et~al.},
\newblock \emph{Appl. Phys. Lett.} \textbf{90}, 164101 (2007) 

\bibitem{lopez2008PRL}
C.~E. L{\'{o}}pez, G.~Romero, F.~Lastra, E.~Solano, J.~C. Retamal,
\newblock \emph{Phys. Rev. Lett.} \textbf{101}, 080503 (2008) 

\bibitem{bai2009PRA}
Y.-K. Bai, M.-Y. Ye, Z.~D. Wang,
\newblock \emph{Phys. Rev. A} \textbf{80}, 044301 (2009) 

\bibitem{Lopez2010PRA}
C.~E. L\'{o}pez, G.~Romero, J.~C. Retamal,
\newblock \emph{Phys. Rev. A} \textbf{81}, 062114 (2010) 

\bibitem{horodecki2001PRA}
K.~Zyczkowski, P.~Horodecki, M.~Horodecki, R.~Horodecki,
\newblock \emph{Phys. Rev. A} \textbf{65}, 012101 (2001) 

\bibitem{tanas}
Z.~Ficek, R.~Tana\'{s},
\newblock \emph{Phys. Rev. A} \textbf{74}, 024304 (2006) 

\bibitem{palermocatania2010PRA}
B.~Bellomo, G.~Compagno, A.~D'Arrigo, G.~Falci, R.~{Lo Franco}, E.~Paladino,
\newblock \emph{Phys. Rev. A} \textbf{81}, 062309 (2010) 

\bibitem{palermocatania2011IJQI}
B.~Bellomo, G.~Compagno, A.~D'Arrigo, G.~Falci, R.~{Lo Franco}, E.~Paladino,
\newblock \emph{Int. J. Quant. Inf.} \textbf{9}, 63 (2011) 

\bibitem{Dijkstra2010PRL}
A.~G. Dijkstra, Y.~Tanimura,
\newblock \emph{Phys. Rev. Lett.} \textbf{104}, 250401 (2010) 

\bibitem{Man2010NJP}
Z.-X. Man, Y.-J. Xia, N.~B. An,
\newblock \emph{New J. Phys.} \textbf{12}, 033020 (2010) 

\bibitem{horodecki1995PLA}
M.~Horodecki, P.~Horodecki, R.~Horodecki,
\newblock \emph{Phys. Lett. A} \textbf{200}, 340 (1995) 

\bibitem{bellomo2010PLA}
B.~Bellomo, R.~{Lo Franco}, G.~Compagno,
\newblock \emph{Phys. Lett. A} \textbf{374}, 3007 (2010) 

\bibitem{Paris2011PRA}
S.~Cialdi, D.~Brivio, E.~Tesio, M.~G.~A. Paris,
\newblock \emph{Phys. Rev. A} \textbf{83}, 042308 (2011) 

\bibitem{harkonen2009PRA}
K.~Harkonen, F.~Plastina, S.~Maniscalco,
\newblock \emph{Phys. Rev. A} \textbf{80}, 033841 (2009) 

\bibitem{francica2009PRA}
F.~Francica, S.~Maniscalco, J.~Piilo, F.~Plastina, K.-A. Suominen,
\newblock \emph{Phys. Rev. A} \textbf{79}, 032310 (2009) 

\bibitem{francica2010PRA}
F.~Francica, F.~Plastina, S.~Maniscalco,
\newblock \emph{Phys. Rev. A} \textbf{82}, 052118 (2010) 

\bibitem{mazzola2010JPB}
L.~Mazzola, S.~Maniscalco, J.~Piilo, K.-A. Suominen,
\newblock \emph{J. Phys. B, At. Mol. Opt. Phys.} \textbf{43}, 085505 (2010) 

\bibitem{viola1998PRA}
L.~Viola, S.~Lloyd,
\newblock \emph{Phys. Rev. A} \textbf{585}, 040101(R) (1998) 

\bibitem{vandersypen2005RMP}
L.~M.~K. Vandersypen, I.~L. Chuang,
\newblock \emph{Rev. Mod. Phys.} \textbf{76}, 1037 (2005) 

\bibitem{facchi2001PRL}
P.~Facchi, H.~Nakazato, S.~Pascazio,
\newblock \emph{Phys. Rev. Lett.} \textbf{86}, 2699 (2001) 

\bibitem{muhktar2010PRA1}
M.~Mukhtar, T.~B. Saw, W.~T. Soh, J.~Gong,
\newblock \emph{Phys. Rev. A} \textbf{81}, 012331 (2010) 

\bibitem{yablonovitch1987PRL}
E.~Yablonovitch,
\newblock \emph{Phys. Rev. Lett.} \textbf{58}, 2059 (1987) 

\bibitem{john1990PRL}
S.~John, J.~Wang,
\newblock \emph{Phys. Rev. Lett.} \textbf{64}, 2418 (1990) 

\bibitem{john1994PRA}
S.~John, T.~Quang,
\newblock \emph{Phys. Rev. A} \textbf{50}, 1764 (1994) 

\bibitem{lodhal2004Nature}
P.~Lodahl, \emph{et~al.},
\newblock \emph{Nature} \textbf{430}, 654 (2004) 

\bibitem{gaoxiang2009PRA}
M.~Al-Amri, G.-X. Li, R.~Tan, M.~S. Zubairy,
\newblock \emph{Phys. Rev. A} \textbf{80}, 022314 (2009) 

\bibitem{zhang2010EPJD}
Y.~J. Zhang, Z.~X. Man, Y.~J. Xia, G.~C. Guo,
\newblock \emph{Eur. Phys. J. D} \textbf{58}, 397 (2010) 

\bibitem{wang2011OptComm}
J.~Wang, L.~Jiang, H.~Zhang, T.-H. Huang, H.-Z. Zhang,
\newblock \emph{Opt. Comm.} \textbf{284}, 5323 (2011) 

\bibitem{wang2012OptComm}
J.~Wang, \emph{et~al.},
\newblock \emph{Opt. Comm.} \textbf{285}, 2961 (2012) 

\bibitem{mazzolapalermo2010PRA}
L.~Mazzola, B.~Bellomo, R.~{Lo Franco}, G.~Compagno,
\newblock \emph{Phys. Rev. A} \textbf{81}, 052116 (2010) 

\bibitem{werlang2009PRA}
T.~Werlang, S.~Souza, F.~F. Fanchini, C.~J.~V. Boas,
\newblock \emph{Phys. Rev. A} \textbf{80}, 024103 (2009) 

\bibitem{mazzola2011frozen}
L.~Mazzola, J.~Piilo, S.~Maniscalco,
\newblock \emph{Int. J. Quant. Inf.} \textbf{9}, 981 (2011) 

\bibitem{Modi2011arxiv}
K.~Modi, A.~Brodutch, H.~Cable, T.~Paterek, V.~Vedral,
\newblock arXiv:1112.6238

\bibitem{mazzola2010PRL}
L.~Mazzola, J.~Piilo, S.~Maniscalco,
\newblock \emph{Phys. Rev. Lett.} \textbf{104}, 200401 (2010) 

\bibitem{can-guo2010NatComm}
J.-S. Xu, \emph{et~al.},
\newblock \emph{Nat. Commun.} \textbf{1}, 7 (2010) 

\bibitem{xu2010PRA}
J.-S. Xu, \emph{et~al.},
\newblock \emph{Phys. Rev. A} \textbf{82}, 042328 (2010) 

\bibitem{haikka2012arxiv}
P.~Haikka, and S.~Maniscalco,
\newblock arXiv:1203.6469

\bibitem{zhou2010QIP}
D.~Zhou, A.~Lang, R.~Joynt,
\newblock \emph{Quantum Inf. Process.} \textbf{9}, 727 (2010) 

\bibitem{lofranco2012PhysScripta}
R.~{Lo Franco}, A.~{D'Arrigo}, G.~Falci, G.~Compagno, E.~Paladino,
\newblock \emph{Phys. Scripta} \textbf{T147}, 014019 (2012) 

\bibitem{bellomo2012PhysScripErika}
B.~Bellomo, R.~{Lo Franco}, E.~Andersson, J.~D. Cresser, G.~Compagno,
\newblock \emph{Phys. Scripta} \textbf{T147}, 014004 (2012) 

\bibitem{lofranco2012PRA}
R.~{Lo Franco}, B.~Bellomo, E.~Andersson, G.~Compagno,
\newblock \emph{Phys. Rev. A} \textbf{85}, 032318 (2012) 

\bibitem{alickibook}
R.~Alicki, K.~Lendi,
\newblock \emph{Quantum Dynamical Semigroups and Applications}
\newblock (Lect. Notes Phys. 717. Springer, Berlin Heidelberg, 2007)

\bibitem{plenio2010PRL}
{\'{A}}.~Rivas, S.~F. Huelga, M.~B. Plenio,
\newblock \emph{Phys. Rev. Lett.} \textbf{105}, 050403 (2010) 

\end{thebibliography}
\end{document}